\documentclass[twoside]{article}
\usepackage{fleqn,espcrc2}

\newcommand{\AmS}{{\protect\the\textfont2
  A\kern-.1667em\lower.5ex\hbox{M}\kern-.125emS}}
 \newcommand{\be}{\begin{equation}}\newcommand{\ee}{\end{equation}}
\newcommand{\bea}{\begin{eqnarray}}\newcommand{\eea}{\end{eqnarray}}
\newcommand{\nn}{\nonumber}\newcommand{\p}[1]{(\ref{#1})}
\newcommand{\lb}{\label}

\newcommand{\cA}{{\cal A}}
\newcommand{\bcA}{\bar{\cal A}}

\newcommand\T{\mbox{Tr}\;}

\newcommand\Y{{\s Y}}

\newcommand\s{\scriptscriptstyle}
\newcommand\q{\quad}

\newcommand{\pp}{{=\!\!\!|}}
\newcommand{\xp}{x^\pp}
\newcommand{\xm}{x^=}

 \newcommand{\Pp}{\partial_\pp}
\newcommand{\Pm}{\partial_=}

\newcommand{\PY}{\partial_\Y}
\newcommand{\bPY}{\bar\partial_\Y}

\newcommand{\tpi}{\theta^+_i}

\newcommand{\tmi}{\theta^-_i}

\newcommand{\btpi}{\bar{\theta}^{i+}}

\newcommand{\btmi}{\bar{\theta}^{i-}}

\newcommand{\bppt}{\bar{\partial}_{2+}}

\newcommand{\bpmt}{\bar{\partial}_{2-}}

\newcommand{\bpmf}{\bar{\partial}_{4-}}
\newcommand{\bppf}{\bar{\partial}_{4+}}
\newcommand{\ppo}{\partial^1_+}

\newcommand{\pmo}{\partial^1_-}

\newcommand{\pmh}{\partial^3_-}
\newcommand{\pph}{\partial^3_+}

\newcommand{\tpo}{\theta^+_1}
\newcommand{\tpt}{\theta^+_2}
\newcommand{\tmo}{\theta^-_1}
\newcommand{\tmt}{\theta^-_2}
\newcommand{\tmh}{\theta^-_3}

\newcommand{\tmf}{\theta^-_4}
\newcommand{\tpf}{\theta^+_4}
\newcommand{\btpo}{\bar\theta^{1+}}
\newcommand{\btmo}{\bar\theta^{1-}}

\newcommand{\btmt}{\bar\theta^{2-}}
\newcommand{\btph}{\bar\theta^{3+}}
\newcommand{\btmh}{\bar\theta^{3-}}
\newcommand{\btpf}{\bar\theta^{4+}}
\newcommand{\btmf}{\bar\theta^{4-}}
\newcommand{\Dpo}{D_+^1}
\newcommand{\Dmo}{D_-^1}
\newcommand{\Dpt}{D_+^2}

\newcommand{\bDph}{\bar{D}_{3+}}

\newcommand{\bDpf}{\bar{D}_{4+}}
\newcommand{\bDmf}{\bar{D}_{4-}}
\newcommand{\Dot}{D^1_2}
\newcommand{\Dto}{D^2_1}

\newcommand{\Vto}{V^2_1}
\newcommand{\Doh}{D^1_3}

\newcommand{\Dtf}{D^2_4}

\newcommand{\Dhf}{D^3_4}
\newcommand{\Dfh}{D^4_3}

\newcommand{\Vfh}{V^4_3}

\newcommand{\N}{\nabla}
\newcommand{\bN}{\bar\nabla}

\title{ $N=4$  SUPER-YANG-MILLS
EQUATIONS  IN HARMONIC SUPERSPACE}

\author{B.M. Zupnik\address{
 Bogoliuibov Laboratory of Theoretical Physics, Joint Institute
for Nuclear Research\\ Dubna, Moscow Region, 141980, Russia;
 e-mail: zupnik@thsun1.jinr.ru}}

\begin{document}

\begin{abstract}
We analyze the superfield equations of the 4-dimensional
$N{=}4$ SYM-theory using light-cone gauge conditions and the
harmonic-superspace approach. The  harmonic superfield equations of
motion are drastically simplified  in this gauge, in particular, the
basic harmonic-superfield matrices  and the corresponding harmonic
analytic gauge connections become nilpotent on-shell.
\vspace{1pc}
\end{abstract}

\maketitle

\section{Introduction}

It is known that the superfield constraints of the $N=3$ and $N=4$
super-Yang-Mills (SYM) theories \cite{So} are equivalent to the
corresponding equations of motion. Moreover, the  component
fields of these theories coincide on mass shell.
 In the harmonic approach
to the $N=3$ SYM-theory \cite{GIKOS}, the $SU(3)/U(1)\times U(1)$
harmonics have been used for the covariant reduction of the spinor
coordinates and derivatives and for the off-shell description of this
theory in terms of the corresponding G-analytic superfields.
 As it has been shown in Ref.\cite{NZ} the light-cone version of the
$N=3$ harmonic superspace simplifies
drastically  superfield equations of the $N=3$ SYM-theory.

Harmonic superspaces of the $D=4, N=4$ supersymmetry have been considered
in Refs. \cite{IKNO,Ba,HH,AFSZ}, in particular, it has been shown that the
self-duality condition for the $N=4$ superfield strengthes corresponds
to the special reality condition in the harmonic superspace.
Stress that the $N=4$
harmonic superspace describes the on-shell superfields only in contrast
to analogous harmonic formalisms for $N=2$ and $3$.

 The short on-shell harmonic superfields
in the Abelian $N=4$  SYM-theory satisfy the
constraints of chirality or different types of harmonic and Grassmann
analyticities \cite{AFSZ}.  It will be shown that these analyticities
are also useful in the $N=4$ non-Abelian SYM-theory.

We shall analyze the classical solutions of the harmonic-superfield
equations  using the convenient light-cone gauge conditions for superfield
connections (see, e.g. \cite{BLN,DL,GS}). These gauge conditions yield the
nilpotent superfield matrices
in the  bridge representation of the $N=4$ SYM-theory.

\section{Harmonic-superspace formulation of   $N=4$ SYM
equations}

The covariant coordinates of the $D=4,~N=4$ superspace are
\be
z^M=(x^{\alpha\dot\alpha} ,\theta^\alpha_i ,\bar\theta^{i\dot\alpha} )~,
\lb{A2b}
\ee
where $\alpha,~\dot\alpha$ are the $SL(2,C)$ indices
and $i$  are indices of the fundamental
representations of the  group $SU(4)$.

We shall study solutions of the  SYM-equations using the non-covariant
notation
\bea
&\xp\equiv x^{1\dot{1}} =t+x^3~,\q \xm\equiv x^{2\dot{2}}=t-x^3~,&\nn\\
&y\equiv x^{1\dot{2}}=x^1+ix^2~,\q\bar{y}\equiv x^{2\dot{1}} =x^1-ix^2~,&
\nn\\
&(\tpi,~\tmi)\equiv
\theta^\alpha_i~, \q (\btpi,~\btmi)\equiv  \bar\theta^{i\dot\alpha}
~.&\lb{A2}
\eea
suitable when the Lorenz symmetry is reduced to $SO(1,1)$.
The general $N=4$ superspace has the odd dimension (8,8) in this
notation.

 The $D=4,~N=4$ SYM-constraints \cite{So}
have the following reduced-symmetry form:
\bea
&\{\nabla^k_+,\nabla^l_+\}=0~,\q \{\bar{\nabla}_{k+},\bar{\nabla}_{l+}\}
=0~,&\nn\\
& \{\nabla^k_+,\bar{\nabla}_{l+}\}=2i\delta^k_l\nabla_\pp~,&\lb{a8}\\
&\{\nabla^k_+,\nabla^l_-\}=W^{kl}
~,\q\{\nabla^k_+,\bar{\nabla}_{l-}\}=2i\delta^k_l \nabla_y~,&\lb{a9}\\
&\{\nabla^k_-,\bar{\nabla}_{l+}\}{=}2i\delta^k_l\bar{\nabla}_y~,\q
\{\bar{\nabla}_{k+},\bar{\nabla}_{l-}\}{=}W_{kl}~,&\lb{a10}\\
&\{\nabla^k_-,\nabla^l_-\}=0~,\q \{\bar{\nabla}_{k-},\bar{\nabla}_{l-}\}
=0~,&\nn\\
& \{\nabla^k_-,\bar{\nabla}_{l-}\}=2i\delta^k_l\nabla_=&\lb{a11}
\eea
where $\nabla$ are the covariant derivatives in the
$(4|8,8)$-dimensional superspace, $W_{kl}$ and $W^{kl}$ are the
gauge-covariant superfields constructed from the gauge connections. These
superfields satisfy the  subsidiary conditions
\be
W^{ik}\equiv \overline{W_{ik}}=-{1\over2}\varepsilon^{ikjl}W_{jl}~.
\lb{A11b}
\ee

The equations of motion for the superfield strengthes follow from the
Bianchi identities
\bea
&& \nabla^i_\pm W^{kl} + \nabla^k_\pm W^{il}=0~,\nn\\
&& \bar{\nabla}_{i\pm} W^{kl}={1\over2}(\delta^k_i
\bar{\nabla}_{j\pm}W^{jl}- \delta^l_i\bar{\nabla}_{j\pm}
W^{jk})~.\lb{a12}
\eea

Let us consider the light-cone gauge conditions
\be
A^k_+=0~,\q \bar{A}_{k+}=0~,\q A_\pp =0~,\lb{a14}
\ee
then the constrains \p{a8} are solved explicitly.

The harmonic superspaces for the $N=4$ SYM-theory have been discussed
in Refs.\cite{IKNO}-\cite{AFSZ}. It has been shown that the G- and
H-analytic
Abelian on-shell superfield strength  lives in the harmonic
superspace with (4+4) Grassmann coordinates. We shall use the analogy with
the HSS description of the $N=3$ SYM-equations \cite{NZ} and consider
the gauge invariance and geometric structure of the superfield $N=4$
equations. Stress that the variety of different G-analytic superspaces
for $N=4$ is more rich than for the case $N=2$ and 3, however,  we do not
know the off-shell superfield structure of the $N=4$ SYM-theory.

We shall use the $SU(4)/ U(1)^3$ harmonics \cite{IKNO,Ba,AFSZ} for the HSS
interpretation of the non-Abelian $N=4$ constraints (\ref{a8}-\ref{a11})
by analogy with the Abelian case.
These harmonics parametrize the
corresponding coset space. They form an $SU(4)$ matrix and are
defined modulo $U(1)\times U(1)\times U(1)$ transformations
\bea
&u^1_i=u^{(1,0,1)}_i~,\quad u^2_i=u^{(-1,0,1)}_i~,&\nn\\
&u^3_i=u^{(0,1,-1)}_i~,\quad u^4_i=u^{(0,-1,-1)}_i &\lb{I1} \eea
where $i$ is the index of the quartet representation of $SU(4)$.
The complex conjugated harmonics have opposite $U(1)$ charges
\bea
&u_1^i=u^{i(-1,0,-1)}~,\quad u_2^i=u^{i(1,0,-1)}~,\nn\\
&u_3^i=u^{i(0,-1,1)}~,\quad u_4^i=u^{i(0,1,1)}~.&\lb{I2}
\eea
Note
that we use indices $I, J=1, 2, 3, 4$ for the projected components
of the harmonic matrix which do not transform with respect to the
'ordinary' $SU(4)$ transformations. The authors of Ref.\cite{HH} prefer
to use the $SU(4)/S(U(2)\times U(2))$ harmonics for the $N=4$ theory.

The corresponding harmonic derivatives $\partial^I_J$ act on these
harmonics and satisfy the $SU(4)$ algebra.

The special  conjugation of the $SU(4)$ harmonics has the following form:
\bea
&u^1_i\leftrightarrow u^i_4~,\q u^2_i\leftrightarrow u^i_3~,\nn\\
&u^3_i\leftrightarrow u^i_2~,\q u^4_i\leftrightarrow u^i_1&\lb{I14}
\eea
 and the conjugation of the harmonic derivatives is
\be
\partial^1_2 f\leftrightarrow-\partial^3_4\widetilde{f},\q
\partial^1_4 f\leftrightarrow-\partial^1_4\widetilde{f},
\ee
where $f(u)$ is an arbitrary harmonic function.

The analytic coordinates in the $N=4$ superspace $H(4,12|6,6)$
are
\bea
&\zeta{=}(X^\pp,X^=,Y,\bar{Y}| \theta^\pm_2,
\theta^\pm_3,\theta^\pm_4,\bar\theta^{1\pm},\bar\theta^{2\pm},
\bar\theta^{3\pm})&\\
&X^\pp=\xp +i(\tpf\btpf -\tpo\btpo)\;,&\nn\\
&X^= =\xm +
i(\tmf\btmf -\tmo\btmo)\;,&\nn\\
& Y=y+i(\tpf\btmf -\tpo\btmo)\;,&\nn\\
&\bar{Y}=\bar{y}+i(\tmf\btpf -\tmo\btpo)\;,&\nn\\
&\theta^\pm_I=\theta^\pm_k u^k_I~,\q\bar\theta^{I\pm}=
\bar\theta^{\pm k}u_k^I~.&
\lb{I15}
\eea

The spinor derivatives have the following simple form in these
coordinates:
\bea
&&D^1_\pm =\partial^1_\pm~,\q
\bar{D}_{4\pm} =\bar\partial_{4\pm}~,  \lb{I16} \\
&& D^2_+ =\partial^2_+ +i\bar\theta^{2+}\Pp+i\bar\theta^{2-}\PY~,\\
&&  D^2_- =\partial^2_- +i\bar\theta^{2+}\bPY+i\bar\theta^{2-}\Pm~,\\
&&
\bar{D}_{1+} =\bar\partial_{1+} +2i\theta^+_1\Pp+2i\theta^-_1\bPY~,\\
&& \bar{D}_{1-} =\bar\partial_{1-} +2i\theta^+_1\PY+2i\theta^-_1\Pm~.
 \eea

The corresponding harmonic derivatives are
\bea
&\Dot =\partial^1_2
+i\tpt\btpo\Pp+i\tpt\btmo\PY &\nn\\
&+i\tmt\btpo\bPY
+i\tmt\btmo\Pm-\tpt\ppo &\nn\\
&-\tmt\pmo+\btpo\bppt+\btmo\bpmt~,&\\
& \Dhf =\partial^3_4
+i\tpf\btph\Pp+i\tpf\btmh\PY &\nn\\
&+i\tmf\btph\bPY
+i\tmf\btmh\Pm-\tpf\pph &\nn\\
&-\tmf\pmh+\btph\bppf+\btmh\bpmf~. &
\eea
Other projections of the  Grassmann and harmonic derivatives can be
constructed analogously.

Let us consider the harmonic projections of the CB covariant derivatives
and the corresponding  connections
\bea
&\nabla^I_+=u^I_k \nabla^k_+=D^I_+~,&\\
& \bN_{I+}=u^j_I\bN_{j+}
=\bar{D}_{I+}~,&\\
&\nabla^I_-=u^I_k \nabla^k_-=D^I_- +\cA^I_-~,&\\
& \bN_{I-}=u^j_I\bN_{j-}
=\bar{D}_{I-}+\bcA_{I-}~.&
\lb{B7}
\eea

Taking into account these relations we can transform
the CB-constraints (\ref{a8}-\ref{a11}) to the equivalent
(2,2)-dimensional set of the G-integrability relations:
\be
\{\N^1_\pm,\N^1_\pm\}=\{\N^1_\pm,\bN_{4\pm}\}=\{\bN_{4\pm},
\bN_{4\pm}\}=0~.\lb{B8}
\lb{B10}
\ee

Thus, the $N=4$ SYM-geometry preserves the Grassmann (6,6) analyticity.
It can be shown that the covariant (4,4)-analyticity of superfield
strength $u^1_ku^2_kW^{ik}$ follows from the basic (6,6)-analyticity in
the HSS geometric formalism.

Now we shall discuss the solution of the G-integrability
relations
\bea
& \cA^1_{\pm}(v)=e^{-v}\left(D^1_\pm
e^v\right)~,&\\
& \bcA_{4\pm}(v)=e^{-v}
\left(\bar{D}_{4\pm}e^v\right)~,&\lb{B15}
\eea
where $v(z,u)$ is the superfield bridge matrix.

The gauge transformations of the bridge
\be
e^v~\Rightarrow~e^\lambda e^ve^{-\tau}~,\lb{lambda}
\ee
contain the (6,6)-analytic  AB-gauge  parameters $\lambda$
\be
(D^1_\pm, \bar{D}_{4\pm}) \lambda =0\lb{B16}
\ee
and   the harmonic-independent constrained
CB-gauge parameters $\tau$.

Matrix $e^v$ determines a transform of the CB-gauge superfields
to the analytic basis (AB).
The analytic gauge group acts on the harmonic connections in AB
\bea
&&\nabla^I_K=e^vD^I_Ke^{-v}=D^I_K+V^I_K(v)~,\lb{br1}\\
&&\delta V^I_K =D^I_K\lambda +[V^I_K,\lambda]~.
\lb{B16b}
\eea

Our gauge choice $\cA^1_+=\bcA_{4+}=0$ corresponds to the following
partial gauge conditions for the bridge:
\be
(\Dpo, \bDpf) v=0~.\lb{gauge}
\ee

We treat bridge $v$ as the basic on-shell superfield, so the
 SYM-equations of this approach are formulated  for this
superfield
\be
[D^I_K,e^{-v}\Dmo e^v]=[D^I_K,e^{-v}\bDmf e^v]=0~
\lb{Hanal}
\ee
where $I < K$.

The subsidiary condition \p{A11b} is equivalent to the reality
condition  for the harmonic projection of the superfield strength
$u^1_iu^2_kW^{ik}$ \cite{Ba,HH} and corresponds to the following equation
in the bridge representation:
\be
-\Dpt (e^{-v}\Dmo e^v)=\bDph(e^{-v}\bDmf e^v)~.\lb{selfd}
\ee

By analogy with the $N=3$ formalism \cite{NZ} one can choose the following
light-cone  gauge for the $N=4$ bridge:
\be
v=\tmo b^1 +\btmf \bar{b}_4 +\tmo\btmf d^1_4~,\lb{gauge2}
\ee
where the fermionic matrices $b^1, \bar{b}_4$ and the bosonic matrix
$d^1_4$ are the (6,6) analytic superfields. This bridge is nilpotent
\bea
&& v^2=\tmo\btmf
[\bar{b}_4,b^1]\;,\q v^3=0~,\lb{nilp1}\\
&&e^{-v}= I - v+{1\over2}v^2=I-\tmo b^1- \btmf \bar{b}_4\nn\\&&
+\tmo\btmf
({1\over2}[\bar{b}_4,b^1]-d^1_4 )~.\lb{nilp2}
\eea

In the gauge group $SU(n)$, our superfields satisfy the conditions
\bea
&&(b^1)^\dagger=\bar{b}_4~,\q (d^1_4)^\dagger=-d^1_4~,\\
&& \T b^1=\T d^1_4=0~.
\eea

Consider the parametrization of the basic spinor connections
in our gauge
\bea
&&\cA^1_-(v)=b^1-\tmo (b^1)^2+\btmf f^1_4\nn\\
&&+\tmo\btmf[b^1,f^1_4]~,\lb{cbcon1}\\
&&\bcA_{4-}(v)=\bar{b}_4-\btmf (\bar{b}_4)^2+\tmo\bar{f}^1_4\nn\\
&&-\tmo\btmf
[\bar{b}_4,\bar{f}^1_4]~,\lb{cbcon2}
\eea
where the following auxiliary superfields are introduced:
\bea
&&f^1_4=d^1_4-{1\over2}\{b^1,\bar{b}_4\}\,,\nn\\
&&\bar{f}^1_4=-d^1_4-{1\over2}\{b^1,\bar{b}_4\}\,.\lb{auxil}
\eea

The H-analyticity equations
\be
(\Dot, \Doh, \Dtf, \Dhf)\cA^1_-(v)=0
\ee
are equivalent to the following (6,6)-analytic equations:
\bea
&& (\Dot, \Doh) b^1=-(\tmt, \tmh) (b^1)^2\;,\\
&& (\Dtf, \Dhf) b^1=-(\btmt,
\btmh)f^1_4\;,
\lb{Coef1}\\
&& (\Dot, \Doh) f^1_4=(\tmt, \tmh)[f^1_4,b^1]~,\\
&&(\Dtf, \Dhf) f^1_4=0~.
\lb{Coef4}
\eea

We shall discuss below the relations between the matrices $b^1$ and
$\bar{b}_4$ which arise from the transform of the CB-condition
\p{selfd} to the analytic representation.

Remember that the following covariant Grassmann derivatives are flat in
the  AB-representation of the gauge group before the gauge fixing:
\bea
&&e^v\nabla^1_\pm e^{-v}=D^1_\pm\;,\nn\\
&&e^v\bar{\nabla}_{4\pm}e^{-v}=\bar{D}_{4\pm}\;\lb{C1}
\eea

 The harmonic connections in the bridge representations $V^I_K(v)$
\p{br1} satisfy automatically the harmonic
zero-curvature equations
\be
D^I_KV^J_L-D^J_LV^I_K+[V^I_K,V^J_L]=\delta^J_KV^I_L-\delta^I_LV^J_K~.
\ee

Basic SYM-equations \p{Hanal}are equivalent to the dynamical
G-analyticity conditions
\be
(\Dmo, \bDmf)V^I_K(v)=0~,\q I < K~.\lb{Ganal}
\ee

 In  gauge \p{gauge2}, these  equations
give us the following  relations:
\bea
&&V^1_2(v)=\tmt b^1~,\lb{prep1}\\
&& V^1_3(v)=\tmh b^1
~,\lb{prep2}\\
&&V^3_4=(V^1_2)^\dagger=-\btmh \bar{b}_4~,\lb{prep3}\\
&&V^2_4=(V^1_2)^\dagger=-\btmt \bar{b}_4~,\lb{prep4}
\eea
where all connections are nilpotent.
Similar relations have been considered in the harmonic formalism
of the $N=3$ SYM-theory \cite{NZ}.

 One can also construct the
non-analytic harmonic connections
\be
e^v\Dto e^{-v}=V^2_1=-\tmo \Dto b^1~.
\lb{sol1}
\ee

The conjugated harmonic connection depend, respectively, on  matrix
$\bar{b}_4$ only
\be
V^4_3=(V_1^2)^\dagger=-\btmf \Dfh \bar{b}_4
\;.\lb{sol3}
\ee

It is not difficult to show that the harmonic AB-connections
$V^2_1$  satisfies the partial (8,6)-analyticity condition
\be
\bar{D}_{4\pm} \Vto=0\lb{B23}
\ee
and the conjugated connection possesses the  (6,8)-analyticity
\be
D^1_\pm \Vfh=0~.\lb{B24}
\ee

The basic AB-superfield strengthes can be constructed in terms of the
harmonic connections by analogy with the $N=2$ SYM-theory \cite{Zu2}
\bea
&&W^{12}=-\Dpo\Dmo V^2_1
=-\Dpt b^1
\;,\\
&&W_{34}=-\bDpf\bDmf V^4_3=
-\bDph c_4 \;.
\lb{H7}
\eea
They satisfy the non-Abelian G- and H-analyticity conditions
which generalize the shortness conditions for the corresponding
Abelian superfields \cite{AFSZ}.

The reality condition
\be
W^{12}=-W_{34}  \lb{B27}
\ee
 is equivalent to the single linear differential relation
between the matrices  $b^1$ and $\bar{b}_4$ which  can be easily solved
via the following representation with the anti-Hermitian (6,6)-analytic
bosonic matrix $A^{13}$
\bea
&&b^1=\bDph A^{13}~,\\
&& \bar{b}_4=\Dpt A^{13}~,\lb{solut}\\
&&W^{12}\equiv -W_{34}=-\Dpt\bDph A^{13}~.
\eea

Consider the evident relation
\be
(b^1)^2={1\over2}\bDph[A^{13},\bDph A^{13}]~.
\ee

Equations \p{Coef1} generate the following relations for $A^{13}$
\be
(\Dot, \Doh) A^{13}={1\over2}(\tmt, \tmh) [A^{13},\bDph A^{13}] ~.
\lb{Aeq1}
\ee

Thus, the harmonic-superspace representation and  light-cone
gauge conditions simplify significantly the analysis of the $N=4$
SYM-equations. We hope that this representation allows us to construct
the interesting solutions of these equations.

The author is grateful to E. Ivanov, J. Niederle and E. Sokatchev           
for interesting discussions.
This work is  partially supported
by the grants RFBR-99-02-18417,
RFBR-DFG-99-02-04022 and NATO-PST.CLG-974874.

\end{document}